# GHz nanomechanical resonator in an ultraclean suspended graphene p-n junction

Minkyung Jung,[1, 2, a)] Peter Rickhaus,[1,3] Simon Zihlmann,[1] Alexander Eichler,[3] Peter Makk,[1,4] and Christian Schönenberger[1, b)]

[1]Department of Physics, University of Basel, Klingelbergstrasse 82, CH-4056 Basel, Switzerland

[2]DGIST Research Institute, DGIST, Daegu 42988, Korea

[3]Institute for Solid State Physics, ETH Zurich, CH-8093 Zurich, Switzerland

[4]Department of Physics, Budapest University of Technology and Economics and Nanoelectronics Momentum Research Group of the Hungarian Academy of Sciences, Budafoki ut 8, 1111 Budapest, Hungary

**Abstract**

We demonstrate high-frequency mechanical resonators in ballistic graphene p–n junctions. Fully suspended graphene devices with two bottom gates exhibit ballistic bipolar behavior after current annealing. We determine the graphene mass density and built-in tension for different current annealing steps by comparing the measured mechanical resonant response to a simplified membrane model. In a graphene membrane with high built-in tension, but still of macroscopic size with dimensions 3 × 1 μm$^2$, a record resonance frequency of 1.17 GHz is observed after the final current annealing step. We further compare the resonance response measured in the unipolar with the one in the bipolar regime. Remarkably, the resonant signals are strongly enhanced in the bipolar regime.

a) minkyung.jung@dgist.ac.kr

b) Christian.Schoenenberger@unibas.ch

**Introduction**

Owing to the exceptional mechanical properties of graphene, such as high strength, graphene-based nanoelectromechanical systems have stimulated intensive research activities in recent years.[1-6] For example, extremely high quality factors[7] as well as ultrasensitive mass and force sensors[8] have been demonstrated. In addition, the low mass density and the high maximal tension allows for extremely high fundamental resonance frequencies. This makes graphene an excellent candidate for exploring quantum physics, since it is possible to cool the resonator to the quantum mechanical ground state. Recently, bilayer and multilayer of graphene have been successfully coupled to superconducting microwave resonators and optical cavities and the interaction between light and nanomechanical motion via radiation pressure has been observed.[9-14] The coupling strength between cavities and graphene was sufficiently strong to observe cavity backaction cooling. Furthermore, owing to the large tunability of the resonance frequency, strong coupling to other resonators and parametric amplification have been demonstrated in recent works.[15,16]

In previous works, graphene mechanical resonators were operated in the megahertz (MHz) range. A gigahertz (GHz) graphene mechanical resonator has not been demonstrated yet. However, such resonators are needed in order to reach the quantum regime without having to actively cool the resonator by opto-mechanical side-band cooling.[17-18] Furthermore, graphene mechanical resonators reported previously were operated in the unipolar regime where charge is transported by either electrons or holes (n or p regime).

In this work, we demonstrate a GHz mechanical resonator in a ballistic graphene p–n junction. Fully suspended graphene resonators were fabricated with two bottom gates which are used to control the carrier type and density. To increase the quality of the suspended graphene layer, it is current annealed in a vacuum chamber at low temperature.[19,20] It is known that this procedure increases the electron mobility, yielding ballistic graphene devices.[21] It has been suggested that residues that remained from the device fabrication and other potentially charged adsorbates are desorbed while current annealing.[19] Using a frequency modulation (FM) technique, we measured the mechanical resonance response of the graphene p–n junction at different annealing steps. We extract the graphene mass density and built-in tension and confirm that mass is desorbed.

**Results and discussion**

A device schematic and measurement setup are shown in Fig. 1(a). Devices are fabricated by first defining an array of Ti/Au gates on an oxidized Si substrate. The gates are 45 nm thick, 600 nm wide, and spaced at a pitch of 600 nm. A 20 nm layer of MgO is then deposited to prevent an accidental gate leak. After covering the gate array with a nominally 1 μm thick resist layer (LOR 5A, MicroChem Corp.), an exfoliated graphene flake is transferred onto the LOR aligned to the bottom gate array by using a mechanical transfer technique. 50 nm of Pd source-drain contacts are deposited to define ohmic contacts to the graphene layer. Finally, the LOR layer underneath the graphene flake is e-beam exposed and developed in order to suspend the graphene. The fabrication process is outlined in detail in ref. 20 and 22. The device is then mounted on a circuit board which provides integrated radio frequency (RF) and DC line connectors.

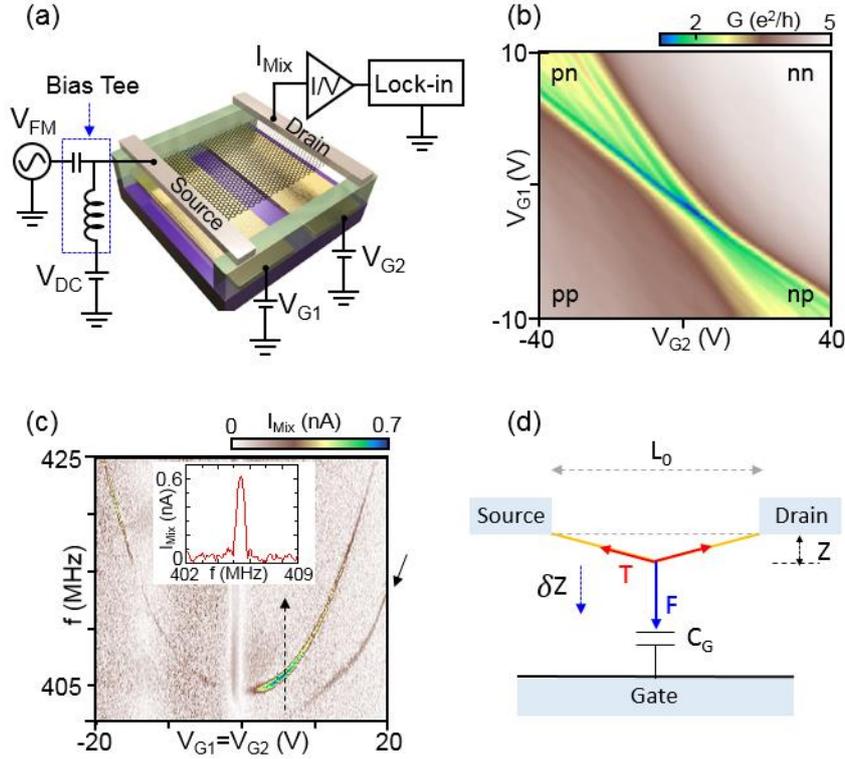

FIG. 1: (a) Schematic of a suspended graphene device with two bottom gates at voltages $V_{G1}$ and $V_{G2}$ and a diagram of the measurement circuit. A frequency-modulated signal $V_{FM}$ with a carrier frequency in the MHz to GHz regime is applied to the source. We measure the mixed-down current $I_{Mix}$ through the graphene by a lock-in amplifier synchronized to the modulation frequency. (b) Differential electrical DC conductance in units of $e^2/h$ as a function of $V_{G1}$ and $V_{G2}$ at $T = 8$ K. Four regions are labeled according to carrier doping, p or n–type, in the left and right graphene regions, controlled by the respective bottom gates. (c) Mixing current $I_{Mix}$ measured as a function of carrier frequency and gate voltage ($V_{G1} = V_{G2}$). Inset: Mixing current vs frequency taken along the dashed line at $V_G$ = 4.8 V, which shows the resonance signal peak. (d) Mechanical model for a graphene resonator simplified to a one-dimensional (1D) string under tension. $L_0$ is the length between the source and drain contacts, $F$ the electrostatic force and $T$ the longitudinal tension in the graphene. $\delta z$ denotes a small time-varying displacement relative to the equilibrium position $z$.

All measurements in this work were carried out in a vacuum chamber with a pressure of typically < $10^{-5}$ mbar at $T \sim 8$ K. We first performed DC conductance measurements of an ultraclean graphene p–n junction using standard lock-in technique. Most as-fabricated devices exhibit a very weak dependence of the conductance on the gate voltage, not showing the suppressed conductance that is expected to arise at the charge-neutrality point (CNP). This is due to strong doping by resist residues. To remove these residues, the device is current-annealed in a vacuum chamber at 8 K until the CNP peak is significantly pronounced.[20] Figure 1(b) shows the differential conductance of device A in units of $e^2/h$ as a function of two bottom gates labeled $V_{G1}$ and $V_{G2}$ by applying a source-drain voltage of $V_{SD}$ = 400 μV after the final current annealing step. The device exhibits four different conductance regions p–p, n–n, p–n, and n–p according to carrier doping in the left and right regions depending on the two bottom gates. In the bipolar region (p–n and n–p) we observe conductance oscillations that can be attributed to Fabry–Pérot interference emerging due to electron waves that interfere with the reflected wave scattered from the p-n junction. The Fabry–Pérot pattern supports that our graphene is in ballistic regime. [21,23]

The resonance of a vibrating graphene device at high-frequencies can be best detected by a mixing method.[24] When applying a small time-varying bias voltage $\delta V(t) = V_{AC} \cos(2\pi f t)$ to the source, keeping the drain contact on ground, the current through the graphene device contains both a linear term $\delta I = G \delta V$ and a squared term $\delta I \propto \delta V^2$, where $G$ is the conductance of the graphene device. The squared term is due to the bias, which causes the total charge in the device to be modulated, modulating $G$ too. Hence, one can write $\delta I = (G + \delta G) \delta V$, where $\delta G \propto \delta V$. The term proportional to $\delta V^2$ mixes a high-frequency signal down to a DC signal.

The applied AC signal also exerts a time-varying force to the graphene membrane, driving a displacement in position indicated in Fig. 1(d) by $\delta z$. This change in displacement leads additionally to a change in gate charge, and hence to a change in conductance $\delta G$. If the AC signal has a frequency $f$ close to the resonance frequency of the membrane $f_0$, $\delta z$ will increase as will $\delta G$. If a FM modulation technique is applied, in which the carrier frequency is modulated with frequency $f_L$, the mixing current can be detected with a conventional lock-in technique synchronized to $f_L$.[7,25] The modulation of the mixing current appears due to the dependence of the vibration amplitude on frequency. A circuit diagram of the measurement setup for the FM technique is shown in Fig. 1(a). The source electrode of the device is connected to a DC source ($V_{DC}$) and an RF generator ($V_{AC}$) via a bias tee. The drain contact is directed to an I/V converter, whose signal is detected in a lock-in amplifier. The graphene resonator is actuated electrostatically by applying a frequency modulated signal with an amplitude $V_{AC}$ at the source electrode. The applied signal at the source electrode can be written as,

$$V_{FM}(t) = V_{AC} \cos(2\pi f t + (f_\Delta/f_L) \sin(2\pi f_L t)) , \qquad (1)$$

where $f$ is the carrier frequency, $f_\Delta$ is the frequency deviation, $t$ is time, and $f_L$ denotes the modulation frequency, which we have typically chosen to be 671 Hz. For a unipolar graphene doubly-clamped membrane, the amplitude of the mixing current can be expressed as,[7,25]

$$I_{mix} = \frac{1}{2}\left|\frac{\partial G}{\partial V_G} V_G \frac{1}{C_G} \frac{\partial C_G}{\partial z} V_{AC} f_\Delta \frac{\partial}{\partial f} Re[\delta z(f)]\right|, \qquad (2)$$

where $G$ is the conductance of the graphene device, $C_G(z)$ is the capacitance between the gate electrode and graphene, and $Re[\delta z(f)]$ is the real part of the graphene oscillation amplitude. Equation (2) can be traced back to the mechanical oscillation of the membrane generating an AC contribution in the gate capacitance, which in turn induces an AC gate charge $\delta Q_{AC}$ that modulates the conductance. It is based on the assumption that $\partial G/\partial Q_{AC}$ is proportional to the transconductance $\partial G/\partial V_G$.[25] Since the dependence of $Re[\delta z(f)]$ changes sign at the resonance frequency, the derivative $\frac{\partial}{\partial f} Re[\delta z(f)]$ has a peak at resonance as shown in the inset of Fig. 1(c). We note also, that $I_{mix} = 0$ for $V_G = 0$, which again is nicely seen in Fig. 1(c).

Figure 1(c) shows a typical resonant response measured after the final current annealing to remove resist residues. The mixing current $I_{Mix}$ is measured as a function of the two bottom gates $V_{G1} = V_{G2}$ and frequency $f$ for a monolayer graphene resonator (Device A, Width/Length $W/L$ = 4.1 μm/1.1 μm) at $T$ = 8 K. The graphene resonant frequency shifts upwards as the gate voltage $|V_{G1=G2}|$ increases due to the tension induced by the gate voltage. The Inset shows a line trace taken along the dashed arrow at $V_{G1=G2}$ = 4.8 V in Fig. 1(c). A line shape with a pronounced peak at the mechanical resonance frequency of approximately 405 MHz is observed. We determine the mechanical quality factor $Q$ to be 600 from the resonance line shape.[7,25] As shown in Fig. 1(c), we observe an additional resonant response at a slightly lower frequency marked by the solid black arrow mostly pronounced in the n–regime. This could be another flexural mode of the resonator, something that we also see in other devices (see e.g. Fig. 2(c)).

To interpret the experimental data, we model the behavior of the graphene resonators. We assume pure uniaxial strain which allows us to apply a simplified model, shown in Fig. 1(d). In this model the membrane is reduced to a 1D string with length $L$. We also assume that there is only a single homogenous gate instead of two gates like in our device. Furthermore, it is assumed that the force of the gate voltage is only acting at the middle of the string. As a result, the resonant frequency $f$ of the graphene membrane can be approximated as (see ESI for the derivation of the equation),

$$f = \frac{1}{2\pi}\sqrt{\frac{1}{\rho wL}\left(\frac{4T_0}{L} + \frac{3}{16}E\left(\frac{C_G' V_G^2}{T_0}\right)^2 - \frac{1}{2}C_G'' V_G^2\right)}, \qquad (3)$$

where $L$ is the length of the suspended graphene membrane, $w$ the width, $\rho$ the 2D mass density, $T_0$ the built-in tension in units of Newton (N), and $E$ the 1D Young's modulus in units of force/meter (N/m). Note, that the built-in tension is defined as $T_0 = T(F)|_{F=0}$, see Fig. 1(d). The primes on $C_G$ denote derivatives with respect to $z$. From the argument in the root we see that the negative term $\propto V_G^2$ dominates if $T_0$ is large, resulting in a prominent negative dispersion of the resonant frequency. As the gate voltage increases, the $V_G^4$ term becomes dominant, leading to an upturn in the dispersion relation. On the other hands, if $T_0$ is small, the $V_G^4$ term dominates also for small values and the resonant frequency shows a positive dispersion with $V_G$ from the beginning. By fitting the experimental data to this model, we aim to determine both $\rho$ and $T_0$ of our devices. To do so, we use in the following $E = 335\ N/m$, for the 2D Young's modulus.[3] This value is deduced from the graphite 3D modulus of 1 TPa, using for the graphene interlayer distance the value 0.335 nm. Details on the fitting procedure are given in the ESI.

We here investigate the evolution of the resonant response as a function of current annealing steps. We could not observe the CNP and resonant responses in the as-fabricated devices due to strong chemical doping by resist residues. As shown in Fig. 2(a), after the initial current annealing the device shows multiple conductance minima, indicating that the graphene sheet is not sufficiently clean. After further current annealing, the device shows a clean CNP and Fabry-Pérot resonances appear in the conductance as shown in Fig. 2(b) (the full plot of conductance is shown in Fig. 1(b)), confirming that the graphene sheet is clean and in the ballistic regime. The mechanical resonant responses $f_0$ for each current annealing step are measured as a function of $V_{G1=G2}$ and displayed in Fig. 2(c) and (d) for the initial and final annealing steps, respectively.

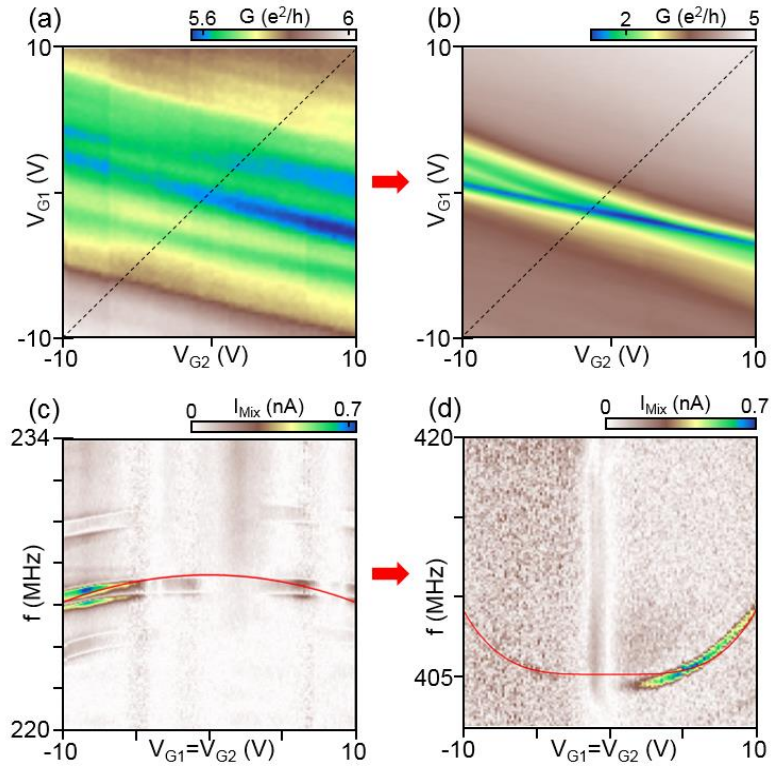

FIG. 2: Differential conductance as a function of $V_{G1}$ and $V_{G2}$ for Device A ($W/L$ = 4.1 μm /1.1 μm) after (a) initial and (b) final current annealing steps. After the final current annealing, the Dirac peak is significantly narrower and pronounced at $V_{G1} = V_{G2} \sim 0$ V, indicating that resist residues are removed. Corresponding mechanical resonant responses for the initial and final current annealing steps are displayed in (c) and (d), respectively. The mechanical mixing current $I_{Mix}$ is measured as a function of frequency $f$ and equal voltage of the two gates (measured along the dashed lines in panel a and b). The applied RF power at the device is $P = -26$ dBm. The solid red lines in (c) and (d) are fits to the equation of a membrane model with the effective graphene mass and built-in tension described in the text.

After the initial current annealing, the resonant frequency shifts downwards with increasing $|V_{G1=G2}|$, as shown in Fig. 2(c). As mentioned above, this indicates that built-in tension is significant. By fitting the data to our membrane model, we estimate the mass density and the built-in tension of the actual membrane to be $\rho = 9.1\rho_0$ and $T_0/W = 4.2$ N/m, respectively, where $\rho_0 = 7.4\times10^{-7}$ kg/m$^2$ is the calculated mass density of monolayer graphene. The estimated mass density is by an order of magnitude larger compared to a clean single layer of graphene, indicating that substantial resist residues still remained on the graphene surface. After the final current annealing, the resonant frequency increased significantly from 226 MHz for the initial annealing to 405 MHz for the final annealing and the frequency shifts upwards as a function of $|V_{G1=G2}|$, indicating that the built-in tension and the mass density decreased significantly. By fitting to Eq. (3) we obtain for the built-in tension $T_0/W = 1.5$ N/m, assuming that the mass density $\rho$ equals the graphene mass density $\rho_0$ when the sample is clean.

The built-in strain values converted from the tensions are estimated to be 1.2 % and 0.4 % after the initial and final current annealing, respectively, showing that the built-in strain significantly decreased after current annealing. This is probably due to the heat in the device generated during current annealing leading to a partial release of the built-in in tension.

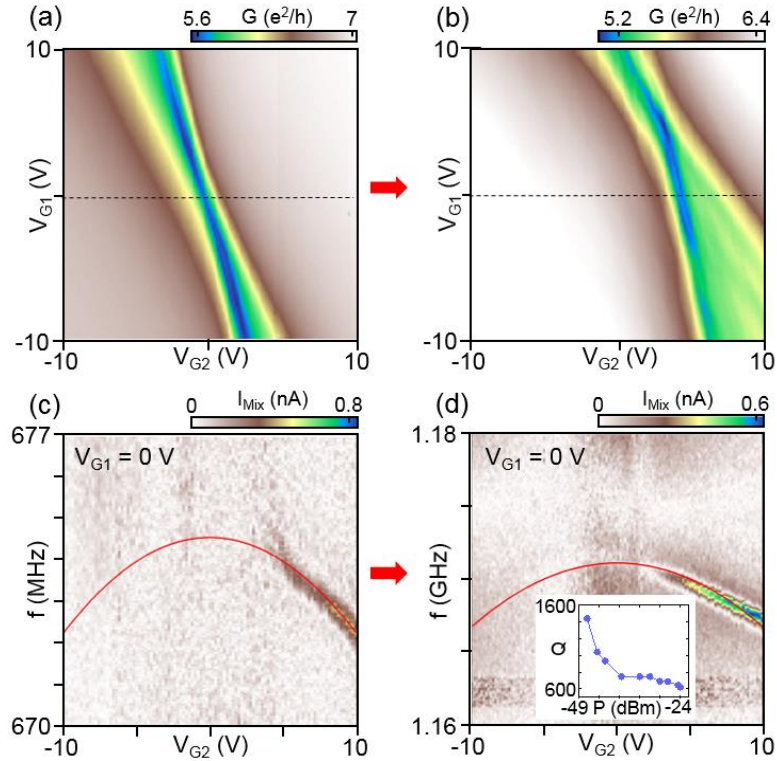

FIG. 3: Differential conductance as a function of $V_{G1}$ and $V_{G2}$ for Device B ($W/L$ = 3 μm /1.2 μm) after (a) initial and (b) final current annealing steps. Corresponding resonant responses are shown in (c) and (d), respectively. The mechanical mixing current $I_{Mix}$ is measured as a function of $f$ and $V_{G2}$ at $V_{G1}$ = 0 V (dashed lines in panels a and b). After the final current annealing step, the resonance frequency increased to ~ 1.17 GHz at $V_{G1}$ = 0 V, which is the highest resonance frequency for a graphene mechanical resonator. The inset of (d) shows the quality factor $Q$ as a function of incident RF power.

Using the same method, we achieved a GHz graphene mechanical resonator in device B. Similar to Fig. 2, we display the conductance map as a function of $V_{G1}$ and $V_{G2}$ for the initial and final current annealing step for device B in Fig. 3(a) and (b), respectively. The corresponding mechanical resonant responses in $I_{Mix}$ as a function of $f$ and $V_{G2}$ at $V_{G1}$ = 0 V for each annealing step are displayed in Fig. 3(c) and (d), respectively. After the final annealing, we observe a remarkably large resonance frequency of $f$ ~ 1.17 GHz. This is the highest mechanical resonance in a fully suspended graphene membrane to our knowledge. Unlike for device A, the frequency as a function of $V_{G2}$ shifts downwards for each annealing step, indicating that the built-in tension is significant, also after the last annealing step. The data fitted to the membrane model confirms that the built-in tension $T_0$ changes only slightly from $T_0/W$ = 16 N/m to $T_0/W$ = 15 N/m, while the mass density is reduced significantly from $\rho$ = 3.3$\rho_0$ to $\rho$ = 1$\rho_0$, increasing the resonance frequency up into the GHz regime. The built-in tension converts to a built-in strain of 4.7 % and 4.4 % for the initial and final current annealing steps, respectively. The large built-in strain of ~ 4 % and the low graphene mass density allows for a mechanical resonance frequency of > 1 GHz for a suspended membrane with μm dimensions. We think that this large built-in tension originates from the device fabrication. The LOR layer on which the graphene membranes is supported deforms as the device is cooled down. This creates a large built-in tension in the device. While the obtained strain of ~ 4 % appears large, comparable values of ~ 2–4 % have been reported before using Raman spectroscopy.[26,27] It is also important to emphasize that the strain value is obtained using a 2D Young's modulus of 340 N/m derived from a 3D Young's modulus of 1 TPa. There have also been reports of larger values for the graphene modulus, which could lower our estimated strain values.[3,28,29]

The inset in Fig. 3(d) shows the extracted quality factor $Q$ as a function of incident RF power. For the lowest power we obtain a $Q \approx 1500$. We then obtain for the quality frequency product $Q \cdot f$ the value $1.8 \times 10^{12}$ s$^{-1}$, which brings this resonator at the measurement temperature of 8 K well into the quantum regime with $Qf > kT/h$, where optical side-band cooling could efficiently be applied to bring the resonator into the ground state.[18] Without additional cooling there are only 150 phonons in this resonator. It is worth noting that a decrease of $Q$ with RF power has been observed before in graphene resonators[7] and might be due to nonlinear coupling between modes.[30,31] It is thus possible that the intrinsic quality factor of our device is significantly higher than what we measured. We further think that higher tension could increase $Q$ through 'dissipation dilution', a technique that has recently enabled ground-breaking nanomechanical devices made from silicon nitride.[32,33]

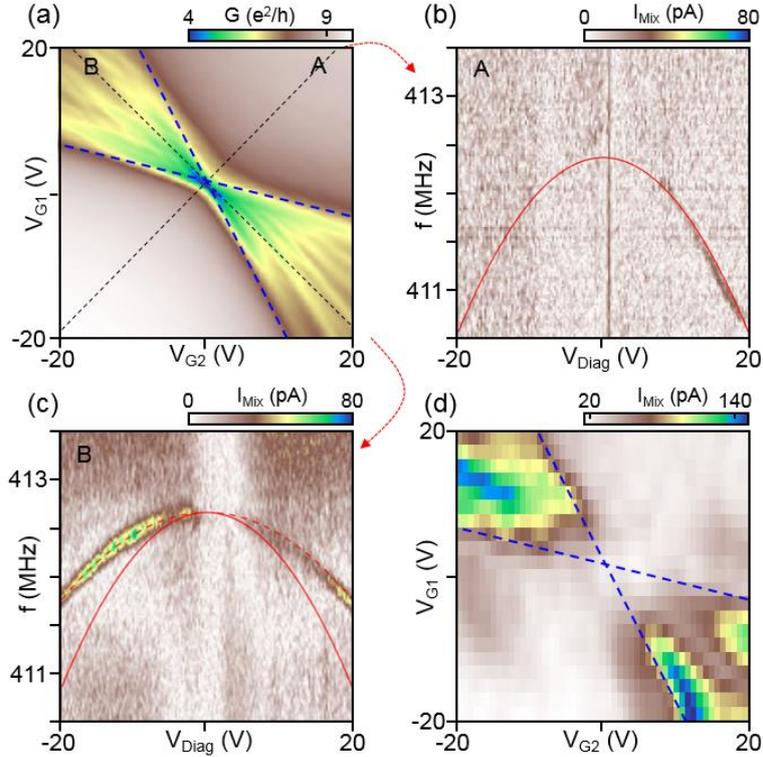

FIG. 4: (a) Differential conductance as a function of $V_{G1}$ and $V_{G2}$ for Device C (W/L = 2 μm /1.5 μm) after final current annealing. (b, c) Resonant responses measured for unipolar (dashed line A) and bipolar gating (dashed line B). The red solid curve in (b) is a fit to the membrane model yielding $\rho = 1\rho_0$ and $T_0/W = 2.8$ N/m. This solid curve is superimposed also in (c) where a bipolar gate voltage is applied. It is seen that the actual frequency shift as a function of $V_{Diag}$ in this experiment is lower. This can be explained by a weaker capacitive coupling due to the built-in p–n junction and indicated with the dashed curve and explained further in the text. (d)

Amplitude of the mixing current $I_{Mix}$ determined from the resonance frequency measurements for each gate voltage. Blue dashed lines indicate the CNPs in the left and right regions and are taken from the conductance measurement in (a). The microwave power during the measurement is kept to $P = -17.5$ dBm. The mixing signal is much stronger in the bipolar regime (n–p or p–n) as compared to the unipolar one (n–n or p–p).

Next, we investigate the amplitude of the mixing current depending on the gate voltage sweep direction in device C. Until now, we only have looked into the unipolar gating condition. Now, we will in addition look into the bipolar case where a p–n junction resides in the graphene device. The conductance as a function of $V_{G1}$ and $V_{G2}$ after the final annealing step is reproduced in Fig. 4(a) and shows distinct conductance values for the unipolar and bipolar condition, as explained in Fig. 1(b). In the bipolar regimes the conductance is markedly lower and Fabry-Pérot resonances are clearly visible, confirming that the graphene sheet is clean. The resonant response measured along the unipolar and bipolar regimes indicated by black dashed lines A and B are displayed in Fig. 4(b) and (c), respectively. The red curve is a fit to the unipolar case in (b) using the membrane model with values $\rho = 1\rho_0$ and $T_0/W = 2.8$ N/m. The frequency response in the bipolar case is markedly smaller. We see that for the same values of gate voltages, the frequency shift in Fig. 4(c) is approximately 50 % of that in (b). The most likely reason for the reduced response in the bipolar case is the charge distribution. Unlike in the unipolar case, where the charge is homogeneously distributed, there is a depletion zone for charge in the middle of the device in the bipolar case. The dynamically added and removed charge appears therefore closer to the source and drain contacts where the graphene membrane is clamped and where it is effectively stiffer. Consequently, the mechanical movement is reduced in the bipolar case, leading to a reduced response of the softening contribution in the frequency dispersion.

For comparison we present the obtained parameters of the three graphene mechanical resonators described in this work in Table 1. It is seen that the resonance frequency $f_0$ systematically increased with annealing, while the built-in tension was reduced. This is only compatible with a reduction of mass. Hence, current annealing does, as anticipated, desorb material, likely resist residues, from the graphene memebrane.

Table 1. Summary of device geometry and parameters of three monolayer graphene mechanical resonators. $W$, $L$ denote the length and width of the graphene membrane, respectively, $f_0$ the zero-applied voltage resonance frequency, the mass density is given relative to the density of a clean monolayer graphene sheet, $T_0/W$ is the zero-voltage pre-tension. In the last column the strain is given obtained by dividing the pre-tension per width with the 2D Young's modulus of graphene denoted by $E_{2D}$.

| Device | $W$ [μm] | $L$ [μm] | Annealing | $f_0$ [MHz] | Mass density ratio [$\rho/\rho_0$] | $T_0/W$ [N/m] | Strain [$T_0/E_{2D}W$, %] |
|---|---|---|---|---|---|---|---|
| A | 4.1 | 1.1 | Initial | 230 | 9.1 | 4.2 | 1.2 |
|   |     |     | Final   | 405 | 1.0 | 1.5 | 0.4 |
| B | 3   | 1.2 | Initial | 674 | 3.3 | 16  | 4.7 |
|   |     |     | Final   | 1170 | 1.0 | 15 | 4.4 |
| C | 2   | 1.5 | Initial | Not measured | | | |
|   |     |     | Final   | 413 | 1.0 | 2.8 | 0.8 |

Interestingly, the mixing signal amplitude scanned along the bipolar regime (dashed line B) is remarkably stronger than that scanned along the unipolar regime (Fig. 4(b) and (c)). A better comparison is provided by Fig. 4(d), displaying the measured mixing amplitude $I_{Mix}$ as a 2D map as a function of gate voltages $V_{G1}$ and $V_{G2}$. The dashed lines mark the CNPs in the left and right regions and are taken from the conductance map in Fig. 4(a). It is seen that the mixing signal in the bipolar region is by up to 10 times larger as compared to the unipolar one. As seen from Eq. (2), the mixing signal is proportional to both the transconductance $dG/dV_G$ and the oscillation amplitude $d(Re[\delta z(f)])/df$ of the resonator. If we assume the same oscillation amplitude at resonance for both the unipolar and bipolar regime, the mixing signal should follow the transconductance. This is what we observe qualitatively when we compare the experiment with numerically calculated transconductance plots (see ESI Fig. S3 for the comparison between the mixing current and calculated transconductance). As can be expected from Fig. 4(a), the transconductance in the bipolar regime is much larger than that in the unipolar one due to the large conductance oscillations induced by Fabry-Pérot interferences. This results in a strong mixing current signal in the bipolar regime. As proven by the data in Fig. 4(d), the bipolar setting in graphene resonators is very convenient for the detection of small mechanical signals, due to the increased sensitivity in this regime.

However, there is also another reason why the mixing current could be substantially larger in the bipolar as compared to the unipolar regime. The photothermoelectric (PTE) effect can become very pronounced in systems with p–n junctions. In our previous work,[34] we observed a strong photocurrent in the bipolar regime of a p–n graphene device when applying a microwave signal, while the photocurrent almost vanished in the unipolar regime. Electron-hole pairs around the CNP cause a temperature gradient in the p–n junction towards the source and drain contacts, generating a photocurrent. The device used in this work has the same structure, so that the microwave used for the mechanical actuation of the graphene sheet can generate a photocurrent as well. There is both an AC and DC (rectified) photocurrent. The former should behave similar to an electrically induced AC current and should therefore yield a mixing current contribution that depends on the mechanical oscillator amplitude. In order to distinguish the two effects, a refined model is needed with which the exact amplitude of the graphene resonator can be calculated, which is beyond the current work.

**Conclusions**

We have demonstrated graphene mechanical resonators with very high frequencies of several 100 MHz to > 1 GHz. We have used ultraclean suspended and current-annealed graphene p–n junctions and determine the graphene mass density and built-in tension after different current annealing steps by fitting the measured resonance frequency dependence on gate voltage to a simplified resonator model. After the final current annealing step, the graphene mass density decreases and likely reaches the pure graphene mass density, indicating that virtually resist residues are removed. In a clean graphene membrane the fundamental mechanic resonance mode has been found to be 1.17 GHz at $V_G = 0$ V. This large resonance frequency for a macroscopic membrane of size 3 × 1.2 μm is only possible due to low mass density of graphene and the high tension that graphene can sustain. In this particular GHz case, the built-in tension is estimated to be $T_0/W \sim 15$ N/m, corresponding to a strain of ~ 4 %. Furthermore, the graphene membrane with two electrically separated gates enables bipolar gating. In this bipolar regime (either p–n or n–p) we have found a strongly enhanced mixing current, while it is weak in the unipolar regime (either n–n or p–p). Our work shows that graphene p–n junctions could be useful for detecting mechanical resonance signals.

**Conflicts of interest**

There are no conflicts to declare.

**Acknowledgements**

CS acknowledges financial support from the ERC projects QUEST & TopSupra and the Swiss National Science Foundation (SNF) through various grants, including the NCCR-QSIT and the Swiss Nanoscience Institute. PM acknowledges funding through OTKA FK-123894 and MJ acknowledges support by the Mid-career Researcher Program (NRF-2017R1A2B4007862) and DGIST R&D Program of the Ministry of Science, ICT, Future Planning (19-NT-01). MJ thanks Hee Chul Park at IBS and Hyeon Jeong Lee, Hwan Soo Jang and Bong Ho Lee at CCRF of DGIST for helpful discussions. PM thanks the support from the Bolayi and Marie Curie Fellowship.